\documentclass[12pt]{article}
\usepackage{amsmath,amssymb,bm,bbm,mathrsfs,amscd}
\usepackage{calc}
\usepackage{color}
\usepackage{amsthm,placeins}
\usepackage{pdfpages,xurl}
\usepackage{hyperref}
\usepackage{graphicx}
\usepackage{dcolumn}
\usepackage{bm}
\renewcommand{\eqref}[1]{Eq.~(\ref{#1})}  
\providecommand{\keywords}[1]
{
  \small	
  \textbf{\textit{Keywords---}} #1
}

\usepackage{pgfplots,subfig}
\pgfplotsset{ 
  compat=newest, 
    colormap={CM}{rgb=(0.,.7,.0) color=(white)},
every tick label/.append style={font=\small},
every axis label/.append style={font=\small}}
\usepgfplotslibrary{fillbetween}

\title{Modelling and predicting the effect of social distancing and travel restrictions on COVID-19 spreading}
\author{Francesco Parino$^{1}$, Lorenzo Zino$^{2}$, Maurizio Porfiri$^3$, Alessandro Rizzo$^{1,4}$}

\date{\normalsize $^1$Department of Electronics and Telecommunications, Politecnico di Torino, 10129 Turin, Italy\\
$^2$Faculty of Science and Engineering, University of Groningen, 9747 AG Groningen, Netherlands\\
$^3$Department of Mechanical and Aerospace Engineering and Department of Biomedical Engineering, New York University Tandon School of Engineering, Brooklyn NY 11201, USA\\
$^4$Office of Innovation, New York University Tandon School of Engineering, Brooklyn NY 11201, USA\\\quad\\
Correspondence should be addressed to: \url{alessandro.rizzo@polito.it}, \url{mporfiri@nyu.edu}}

\begin{document}

\maketitle
\newpage 

\begin{abstract}
To date, the only effective means to respond to the spreading of COVID-19 pandemic are non-pharmaceutical interventions (NPIs), which entail policies to reduce social activity and mobility restrictions. Quantifying their effect is difficult, but it is key to reduce their social and economical consequences. Here, we introduce a meta-population model based on temporal networks, calibrated on the COVID-19 outbreak data in Italy and applied to evaluate the outcomes of these two types of NPIs. Our approach combines the advantages of granular spatial modelling of meta-population models with the ability to realistically describe social contacts via activity-driven networks. We focus on disentangling the impact of these two different types of NPIs: those aiming at reducing individuals' social activity, for instance through lockdowns, and those that enforce mobility  restrictions. We provide a valuable framework to assess the effectiveness of different NPIs, varying with respect to their timing and severity. Results suggest that the effects of mobility restrictions largely depend on the possibility to implement timely NPIs in the early phases of the outbreak, whereas activity reduction policies should be prioritised afterwards. 
\end{abstract}

\section{Introduction}

Following the first report of the novel coronavirus (SARS-CoV-2) in Wuhan, COVID-19 has risen above {$70$ million cases and $1,599,922$ reported deaths as of 13th December 2020} \cite{whoSituation}. The ongoing pandemic quickly reached Europe during February and March 2020, forcing most of the countries to implement unprecedented non-pharmaceutical interventions (NPIs) to fight the spread \cite{prem2020effect, lai2020effect, kraemer2020effect,  tian2020,Haug2020}. Some of these interventions promote policies to reduce human-to-human interactions, for example by enforcing social distancing, halting nonessential activities, closing schools, and banishing large gatherings \cite{prem2020effect,tian2020,Haug2020}. Others limit human mobility by means of travel restrictions and bans \cite{chinazzi2020effect,Haug2020}. Due to the considerable economic and social cost associated with the implementation of both of these types of policies \cite{Bartik2020economic,Bonaccorsi2020economic,Qiu2020}), it is crucial to assess their effectiveness. Mathematical and computational epidemic models are key to accurately evaluate a wide range of what/if scenarios, predicting the evolution of the pandemic for different choices of NPIs \cite{siegenfeld2020opinion,bertozzi2020challenges, gatto2020spread,chinazzi2020effect,aleta2020, metcalf2020mathematical, modelingCovid19,estrada2020covid,dellarossa2020}.

One of the fundamental aspects of the spread of infectious diseases is its spatial {diffusion} and the concurrent role of human mobility patterns \cite{colizza2006role, brockmann2013hidden, balcan2009multiscale}. Extensive studies on mobility within the COVID-19 pandemic revealed that population movements are among the main drivers of the spatial spreading  of the outbreak \cite{kraemer2020effect, jia2020population}. Network structures have emerged as a powerful framework to encapsulate such mobility patterns within mathematical models of epidemics, especially by means of meta-population models \cite{Pastor-Satorras2015}. This modelling paradigm is based on the definition of a set of communities (Provinces, Counties, or Regions), connected by a network that captures daily short-range commuting and long-range mobility.

Different from {most of the classical} meta-population models that {tend to} assume homogeneous mixing within {each community} \cite{Pastor-Satorras2015,Gomes2018}, we propose a network structure that accounts for the inherent, heterogeneous and time-varying nature of human interactions \cite{Volz2009, Holme2012}, together with behavioural changes in response to the pandemic evolution \cite{Funk2010,RizzoPRE2014}. To this aim, individuals interact on the basis of a mechanism inspired by activity-driven networks (ADNs) \cite{Perra2012,Zino2016}. Our model {includes} two key aspects of social communities, mobility patterns and temporal, heterogeneous networks of contacts. Within this meta-population model, we incorporate a variation of a susceptible--infected--removed (SIR) epidemic process \cite{brauer2011mathematical}. Such an epidemic process allows to capture several key features of COVID-19, like the existence of latency periods and the delay in the official reporting of infections and deaths. 

We calibrate the model on epidemic data from the {first wave of the} Italian COVID-19 outbreak \cite{protezioneCivile}, to examine different scenarios that evaluate the spatial effects of NPIs. In particular, we explore the interplay between reduction in social activity and mobility restrictions. At the modelling level, the former mechanism acts upon the network of contacts, while the latter modifies mobility patterns between communities. Our findings reveal that the timing of the interventions is essential toward  effective implementations. We conclude that mobility restrictions should be applied at the early stage of the epidemic and coupled with appropriate policies to reduce social activity. Surprisingly, the impact of mobility restrictions is spatially heterogeneous. For the Italian outbreak, this results in a greater benefit for Southern regions, that is, those located far from the initial outbreak. The overall effect of early travel restrictions in these areas leads to $12\%$ of reduction in the total number of deaths.  We also examine differential interventions among age cohorts, determining that the application of  severe restrictions only to the most vulnerable age cohorts would not be sufficient to effectively reduce the deaths toll. Different phenomena are observed upon the {relaxation} of containment measures, with the contribution of {keeping} mobility restrictions being negligible. In this phase of the fight against the epidemic, policies limiting social activity (for instance, by enforcing the use of face masks or social distancing) {yield} the main benefits in mitigating resurgent outbreaks.

\section{Methods}

\subsection{Model}

\subsubsection{Meta-population activity-driven model}

We consider a population of $n$ individuals partitioned into a set $\mathcal H=\{1,\dots,K\}$ of communities, located in bounded geographical areas (administrative divisions, such as Regions, Provinces or Municipalities), where $n_h$ is the number of individuals in the $h$th community.  Communities are connected through a weighted graph that models travel paths between them. The weight matrix $W\in[0,1]^{K\times K}$, called \emph{routing matrix}, is a matrix {with} non-negative entries, zeros on the main diagonal and row sums equal to 1, such that $W_{hk}$ is the fraction of members of community $h$ that move to community $k$ per unit-time.

\begin{figure}
\centering
\includegraphics[width= 1\textwidth]{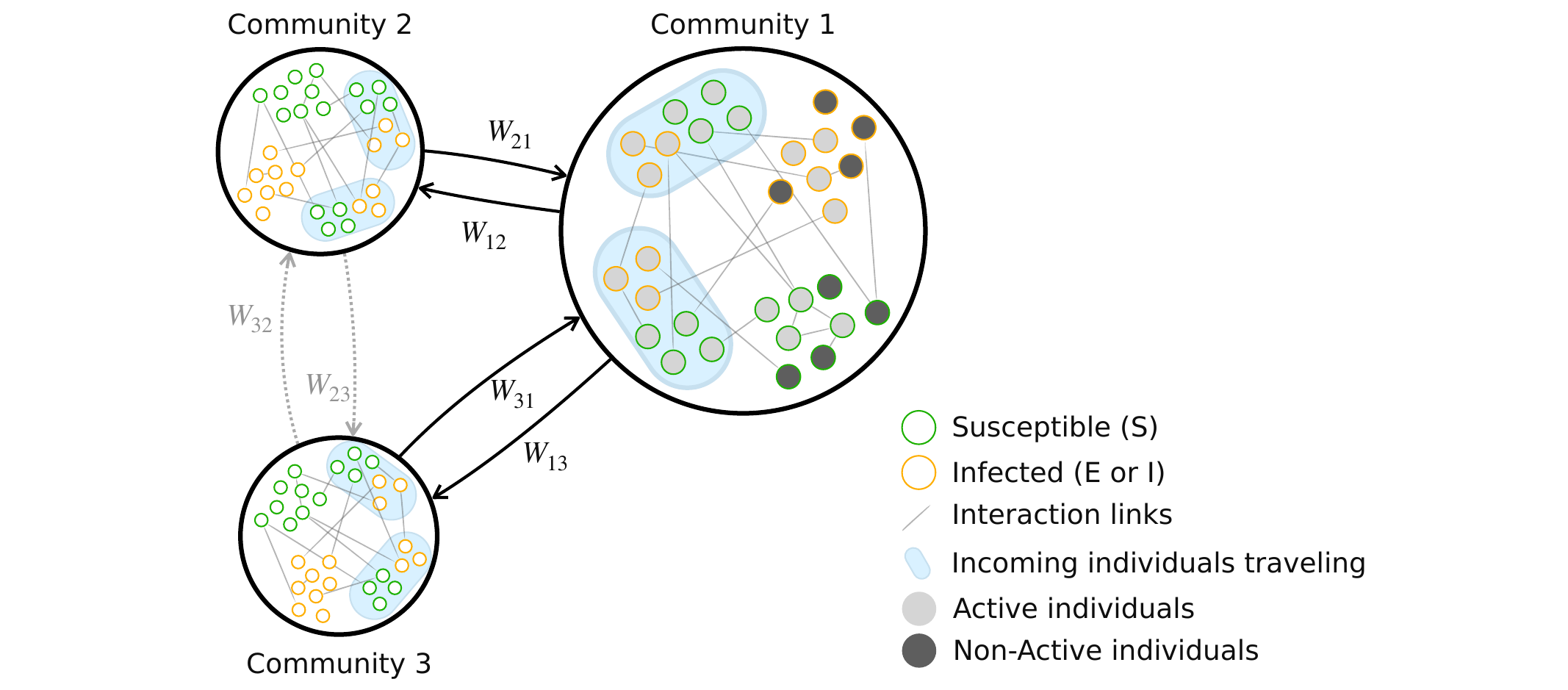}
\caption{Schematics of the meta-population model, {which illustrates the community structure and the role of the routing matrix $W$.}}\label{fig:meta}
\end{figure}

Individuals interact according to a mechanism inspired by ADNs \cite{Perra2012,Zino2016}, which accounts for the inherent, heterogeneous propensity of humans to interact with others. Specifically, individuals are divided into $P$ baseline activity classes $0<a_1<a_2<\dots<a_P\leq 1$, where the \emph{activity}  $a_i$ of individuals in the $i$th class quantifies their nominal propensity to interact with others{, which can be interpreted as the average probability for an individual belonging to the $i$th class to generate interactions in a unit time-step.} At each time-step and for each activity class $i$, a fraction $a_i$ of individuals, {selected uniformly at random,} activates and generates interactions with others, {regardless of their class}. This fraction of the population is called \em active\em. Active individuals may generate interactions within the community where they are located, or they may travel and interact in other communities, {before returning to their community, at the end of the time-step}. This aspect is included in a \emph{mobility parameter} $b\in[0,1]$ that quantifies the baseline fraction of the active population that commutes to other communities; the commuting unfolds according to the routing matrix $W$ (Fig.~\ref{fig:meta}). The remaining fraction $1-b$ of the active population does not commute and interacts locally with individuals randomly selected within their community. We assume that the $P$ activity classes are equally distributed in different communities. Specifically, we introduce the activity distribution vector $\eta\in[0,1]^P$ such that the expected number of individuals with activity class $a_i$ in the $h$th community is equal to the product $n_h\eta_i$. We finally introduce a parameter $m\geq 0$ that captures the {average number of contacts} generated by each active individual in a time-unit.

Two parameters $\alpha\in[0,1]$ and $\beta\in[0,1]$ are introduced to model NPIs. The former, $\alpha$, models individuals' self-isolation  due to the awareness of the disease spreading. In the model, this corresponds to scaling down the individual activity from its baseline value $a_i$ to  $\alpha a_i$. The latter, $\beta$, captures the effect of mobility restrictions, which are modelled by scaling down the baseline mobility parameter from $b$ to $\beta b$.

\subsubsection{Disease progression}

\begin{figure}
\centering
\includegraphics[width= .8\textwidth]{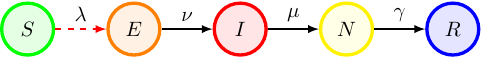}
\caption{Schematics of the epidemic progression. { Individuals may be susceptible to the disease (S), exposed but yet not infectious (E), infectious (I), non-infectious (N), and recovered or dead (R). The compartment $N$ captures the delay between the end of the infectious period and the reporting of a death, and is key for the parameter identification form real-world data. All the transition rates are detailed in the main text and reported in Table~\ref{tab:tabella_parametri}.}}\label{fig:epidemic}
\end{figure}

The disease progression is modelled according to an extension of the classical SIR model (Fig.~\ref{fig:epidemic}), which encapsulates a latency period between contagion and infectiousness, a limited duration of the infectious period, coinciding with the peak of the viral load, and a delay for deaths reporting \cite{prem2020effect}.  Specifically, we adopt a susceptible--exposed--infectious--non-infectious--removed (SEINR) compartmental model (Fig.~\ref{fig:epidemic}).  After contagion, infected individuals become initially exposed ($E$) before spontaneously moving into the infectious ($I$) compartment with rate $\nu$. Once the infectious period terminates (with rate $\mu$), individuals transition to the non-infectious compartment ($N$), before recovering (or dying) with rate $\gamma$, which is represented by the $R$ compartment. The compartment $N$ captures the delay between the end of the infectious period and the reporting of a death. The number of deaths is the most reliable parameter for calibration, given the uncertainty in reporting active infectious cases. The parameters have immediate interpretation: $1/\nu$ is the average latency period of the disease (time from contagion to infectiousness), $1/\mu$ is the average period of communicability (in which individuals are infectious) and  $1/\gamma$ captures the delay before reported deaths. Hence, $1/\mu+1/\gamma$ is the average time from infectiousness to the reported death. {We comment that other important features of COVID-19 may be easily incorporated by considering further compartments into the model, similar to~\cite{giordano2020modelling}. In this vein, one may include, for instance, a differentiation between symptomatic and asymptomatic infectious individuals, which might help implement timely feedback control interventions.}

The contagion mechanism involves an interaction. We introduce a parameter $\lambda\in[0,1]$ that captures the fraction of  susceptible individuals that become exposed after an interaction with an infectious one. The contagion does not depend only on $\lambda$, but also on individual properties (their activity) and network structure, as well as on the prevalence of infectious individuals. We denote by $\Pi_i^h(t)$ the \emph{contagion function}, that is, the fraction of susceptible individuals with activity $a_i$ and who belongs to community $h$ that becomes infected at time $t$, whose expression is detailed in the following.

\subsubsection{Dynamics}

We consider the generic activity class $i$ and community $h\in\mathcal H$. Let $S_i^h$, $E_i^h$, $I_i^h$ and $N_i^h$ be {macroscopic variables counting} the number of susceptible, exposed, infectious and non-infectious individuals of class $i$ in community $h$, respectively. Clearly, $n_h\eta_i-S_i^h-E_i^h-I_i^h-N_i^h$ is the number of removed individuals in class $i$ and community $h$. In the thermodynamic limit of large populations \cite{Pastor-Satorras2015,Kurtz1970,Kurtz1971}, $n\to\infty$, we {describe the epidemic spreading in terms of the macroscopic variables by writing} the following system of {mean-field recurrence} equations:
\begin{equation}\label{eq:ode}\begin{array}{lll}
S_i^h(t+1)&=&(1-\Pi_i^h(t))S_i^h(t)\\
E_i^h(t+1)&=&\Pi_i^hS_i^h(t)+(1-\nu) E_i^h(t)\\
I_i^h(t+1)&=& \nu E_i^h(t)+(1-\mu) I_i^h(t)\\
N_i^h(t+1)&=&\mu I_i^h(t)+(1-\gamma) N_i^h(t).
\end{array}
\end{equation}

We now detail the contagion mechanism and derive an explicit expression for $\Pi_i^h(t)$. To simplify the notation, we omit time $t$, that is, $\Pi_i^h$ is used to denote $\Pi_i^h(t)$. In the thermodynamic limit of large populations $n\to\infty$ and assuming that the epidemic prevalence is small so that we can neglect the probability of having multiple interactions with infectious individuals at the same time, the quantity $\Pi_i^h$ can be written as the sum of four different terms. The first summand accounts for the contagions caused by the fraction $\alpha a_i(1-\beta b)$ of active individuals from $S_i^h$ that remains in the $h$th community and interacts there with infectious individuals; the second summand accounts for the infections caused by the fraction $(1-\alpha \beta a_i b)$ of $S_i^h$ that remains in community $h$ and comes in contact there with active infected individuals; the third and the fourth summands account for the contagions of the fraction $\alpha \beta a_i b$ of $S_i^h$ that is active and moves to other communities interacting with infected individuals or receiving interactions from active infectious individuals in the community they move to, respectively. These four terms yield
\begin{equation}\label{eq:Pi}\begin{array}{lll}
 \Pi_i^h&=&\displaystyle m\alpha a_i(1-\beta b)\lambda  P_h+m(1-\alpha \beta  a_i b)\lambda  Q_h\\&&\displaystyle+m\alpha\beta a_ib\sum_{k\in\mathcal H}W_{hk}\lambda  P_k+m\alpha\beta a_ib\sum_{k\in\mathcal H}W_{hk}\lambda  Q_k,
\end{array}
\end{equation}
where 
\begin{subequations}\begin{align}
P_h=\frac{1}{\tilde n_h}\Bigg(\sum_{j=1}^{P}(1-\alpha\beta  a_j b)I_j^h + \displaystyle\sum_{k\in\mathcal H}W_{kh}\sum_{j=1}^{P}\alpha \beta a_j b I_j^k\Bigg),\\
Q_h=\frac{1}{\tilde n_h}\Bigg(\sum_{j=1}^{P}(1-\beta b)\alpha a_j I_j^k+ \displaystyle\sum_{k\in\mathcal H}W_{kh}\sum_{j=1}^{P}\alpha\beta a_j b I_j^k\Bigg),
\end{align}\end{subequations}
are the fraction of infectious and active infectious individuals that are present in community $h$, respectively. The quantity $$\tilde n_h=(1-\alpha\beta\langle a\rangle b)n_h+\alpha\beta\langle a\rangle b\sum_{k\in \mathcal H}W_{kh}n_k$$ is the number of individuals who are located in community $h$, where  $\langle a\rangle:=\sum_{i=1}^P\eta_i a_i$ is the average activity of the population.

\subsection{Model calibration}

We calibrate the model to reproduce the COVID-19 outbreak in Italy, setting Provinces as communities, using epidemiological parameters from the medical literature \cite{zhang2020evolving, world2020report, zou2020sars}, mobility data by the Italian National Institute of Statistics (ISTAT) \cite{Istat}, and data of officially reported deaths \cite{protezioneCivile}. Based on available empirical data on social contacts per age groups \cite{mossong2008social}, we partition the population in two activity classes. The population below 65 years old forms the \textit{high activity class}, and the population above $65$ years old constitutes the \textit{low activity class}. Different mortality rates are associated with the two classes to estimate the number of deaths, base on serology-informed data \cite{PEREZSAEZ2020}. The details of the model calibration can be found in the following.

\subsubsection{Calibration of the meta-population model}

The Italian territory is divided into $K=107$ Provinces, which are chosen as  communities, extracting the corresponding population $n_h$ from the census data \cite{Istat}. {Provinces are administrative units that offer most of the essential services to the population (supermarkets, hospitals, schools, public offices, restaurants, etc.). Hence, our choice of granularity allows one to distinguish and disentangle the effects of the two categories of NPIs considered in this paper, whereby activity reduction refers to the execution of everyday-life activities within a Province, and mobility restrictions prevent travels between Provinces.} Provinces are grouped in $20$ Regions, gathered in $5$ macro-regions: \emph{North-West}, \emph{North-East}, \emph{Centre}, \emph{South} and \emph{Islands} (Supplementary material, Sec.~S1 and Fig.~S1). 

We partition the population into $P=2$ activity classes, based on age-stratified data on social contacts \cite{mossong2008social}, aggregating age-groups that form high or low number of contacts, respectively. Specifically, the former contains people below 65 years old, while the latter gathers people above 65 years old. According to the same study, we set $m=19.77$. The baseline activity classes $a_1$ and $a_2$ are determined by matching the average number of contacts of the individuals in the classes. The fraction $\eta_i$ of population in each class is determined from the Italian age distribution \cite{Istat}. {Simulations are presented in the Supplementary material (Fig.~S8) to show robustness of our results for different class partitioning.}

\begin{figure}
\includegraphics[width=1\textwidth]{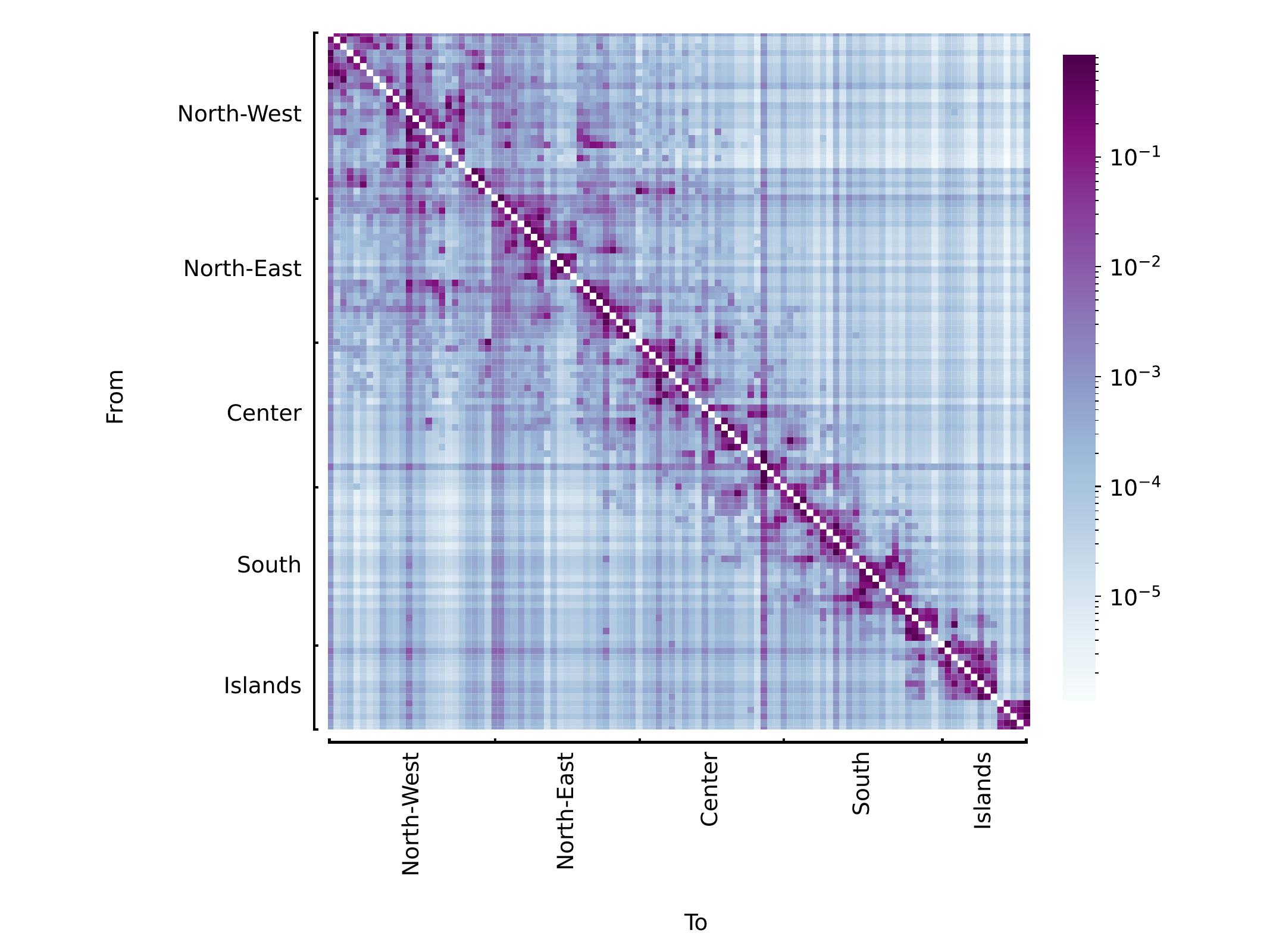}
\caption{Heat-map representing the routing matrix $W$ between Provinces estimated from \cite{Istat}. Colour-code represents the fraction of active people that travel from a Province to another. Provinces are gathered in macro-regions{, as detailed and illustrated in the Supplementary material (Sec.~S1 and Fig.~S1).}}\label{fig:heat}
\end{figure}

We consider two types of mobility: commuting pattern between Provinces and long-range mobility. The former is directly obtained from the $2011$ census data in the ISTAT database \cite{Istat}, which has been validated and adopted to model mobility in recent works on COVID-19 \cite{gatto2020spread}. 
Comprehensive data on long-range mobility is not available. We estimate it as follows. For each province, we consider the number of nights spent in accommodation facilities over the period from February to May 2011, which represents the destinations of travellers \cite{Istat}. Origins are estimated based on the flows between macro-regions \cite{Istat}. Assuming uniformity within each macro-region, we set the origins proportional to the population of each Province. Finally, $W$ is obtained by combining the two origin-destination matrices (Fig.~\ref{fig:heat}). The mobility parameter $b$ is estimated as the fraction of population that moves outside their Province, using  data from \cite{Istat}.

NPIs are implemented as follows. At $t=t_0$, we set $\alpha=\beta = 1$.  Then, based on empirical data \cite{pepe2020covid}, we consider a linear decrease along $15$ days to reach a value $\alpha_{\text{low}}$. Such a decrease begins on 5th March (day of the enforcement of the first social distancing measures) and ended on 20th March  (when a severe lockdown is enacted). Similar, $\beta$ is reduced to $\beta_{\text{low}}$. {As suggested in \cite{pepe2020covid}, mobility restrictions have not been implemented uniformly county-wide: changes were enacted} on 1st March for macro-regions \emph{North-East}, \emph{North-West} and \emph{Centre}, and on 7th March for \emph{South} and \emph{Islands}. The values of $\alpha_{\text{low}}$ and $\beta_{\text{low}}$ are identified from epidemic data.

\subsubsection{Calibration of the epidemic parameters}

Epidemic parameters are taken from the literature on COVID-19. Specifically, the latency period $1/\nu$  and the infectious period $1/\mu$ are taken from \cite{prem2020effect}, based on clinical estimations from \cite{backer2020incubation,woelfel2020clinical}, respectively; $\gamma$ is the inverse of the difference between the average time from infectiousness to the reported death \cite{Linton2020} and $1/\mu$. The infection probability $\lambda$ depends on the model of social interactions. Hence, we identify it from real-world data.
Table~\ref{tab:tabella_parametri} reports the parameters used in our simulations.

\begin{table}
\centering
\begin{tabular}{llll}
 & Meaning& Value(s) &  Ref./Id.
\\ \hline
$1/\nu$  & latency period & 6.4 \mbox{days}&  \cite{prem2020effect,backer2020incubation}
\\
$1/\mu$  & infectiousness period &5 \mbox{days}&  \cite{prem2020effect,woelfel2020clinical}
\\
$1/\gamma$  &time from infectiousness to reported death& 9.52 \mbox{days}&   \cite{prem2020effect,woelfel2020clinical,Linton2020}  \\
$\lambda$ &per-contact infection probability& 0.042 &  \checkmark
\\
$\eta$  &class distribution & {[}0.768, 0.232{]}& \cite{Istat,mossong2008social}\\
$a$ &baseline activity  & {[}0.149, 0.545{]} \mbox{day}$^{-1}$&  \cite{mossong2008social}
\\
$b$ &mobility parameter & 0.09 & \cite{Istat}  
\\
$\alpha_{\text{low}}$&activity reduction  & 0.176 & \checkmark 
\\ 
$m$ &average number of contacts & 19.77 & \cite{mossong2008social}
\\
$\beta_{\text{low}}$ &mobility reduction& 0 & \checkmark 
\\ 
\end{tabular}
\vspace*{6mm}\caption{Model parameters; check-marked parameters are identified {by fitting real-world data of reported death from \cite{protezioneCivile}. From the table, we derive the following parameters: $\nu=0.156$ day$^{-1}$, $\mu=0.2$ day$^{-1}$ and $\gamma=0.105$ day$^{-1}$.}}
\label{tab:tabella_parametri}
\end{table}

\subsubsection{Parameter identification}
We calibrated our model by fitting the temporal evolution of the reported deaths, during the COVID-19 outbreak in Italy. Data at the Regional level were retrieved from the official Italian \textit{Dipartimento della Protezione Civile}~\cite{protezioneCivile} database. 
This database starts on 24th February, and we had extended it backward in time for 20 days (until 4th February). We filled with zero deaths the section of the database from 4th February to 20th February, and we manually corrected the database to include seven deaths that were not reported therein in the period from 20th February to 23rd February (Supplementary material, Sec.~S2). 
To calibrate the model we focused on the period from 4th February (denoted as $t_0$) to 18th May (denoted as $t_{\text{end}}$), namely until the first relaxation of NPIs. To enhance the reliability of the data we applied a weekly moving average.

Using the SEINR epidemic progression model, we computed the deaths in Province $h$ for activity class $i$ as a fraction of the removed individuals $R_i^h(t)$, according to the class fatality ratio $f_1=0.045\%$ and $f_2=5.6\%$. The latter was inferred from a serology-informed estimate performed on age-stratified data from Geneva, Switzerland~\cite{PEREZSAEZ2020}, scaled on the Italian age-distribution using census data~\cite{Istat}. 
Since we had no access to information about the initial number of exposed $E_i^h(t_0)$ or infected $I_i^h(t_0)$ individuals, such initial conditions needed to be identified. 
For each Province $h$ and activity class $i$, we initialised the number of exposed and infected as a fraction $k_1$ and $k_2$ of the total reported cases at the end of the observation time (24th June) $C^h(t_{\text{end}})$ from the official database \cite{protezioneCivile}:
\begin{equation}\label{eq:initial}
    E_i^h(t_0)=k_1\eta_iC^h(t_{\text{end}})\quad\text{and}\quad  I_i^h(t_0)=k_2\eta_iC^h(t_{\text{end}}),
\end{equation}
where $k_1$ and $k_2$ were identified together with the other parameters.

\begin{figure}
\includegraphics[width=\textwidth]{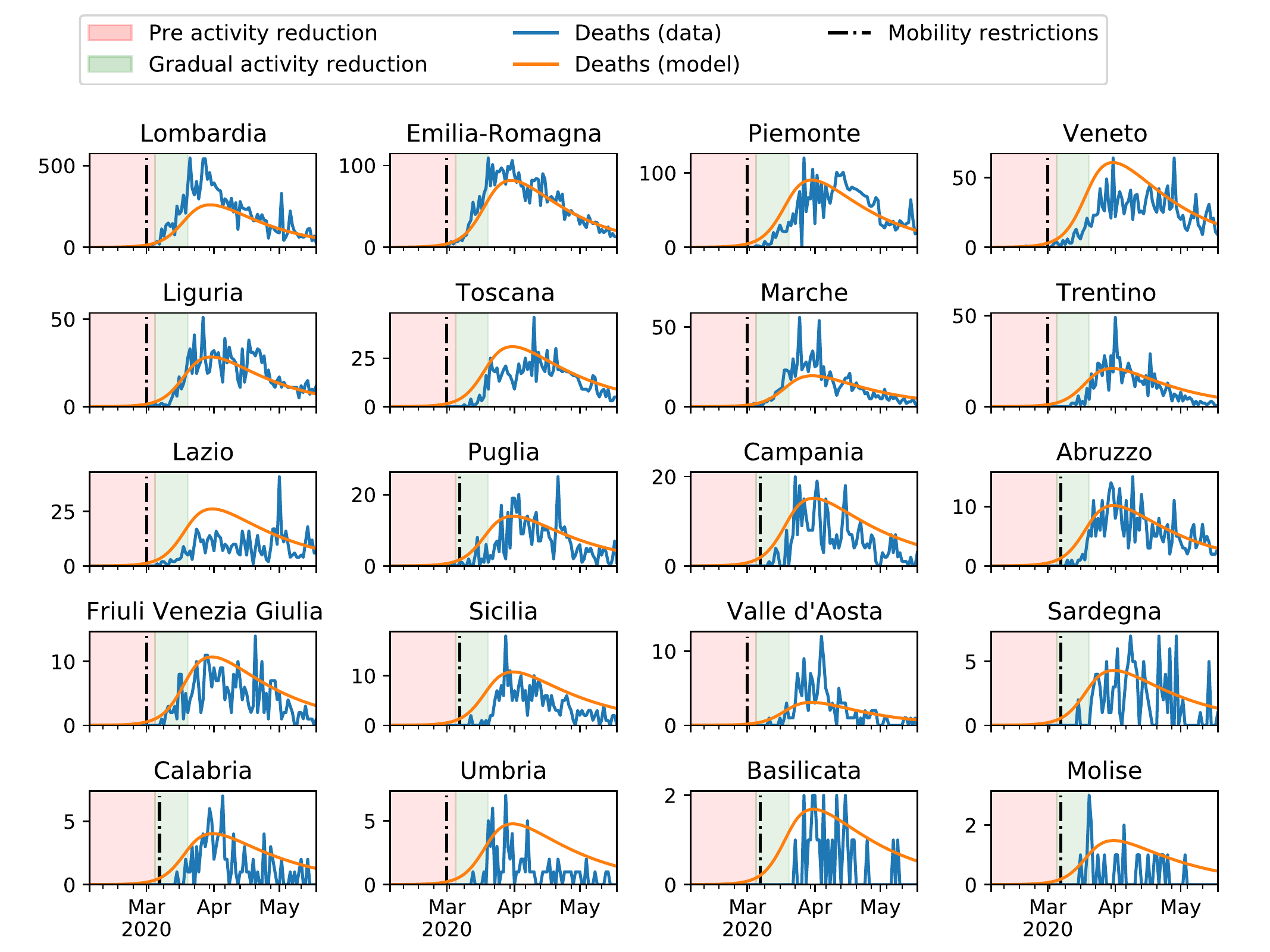}
\caption{Results of the model calibration aggregated at Region level. Model parameters are summarised in Table 1 of the main document. The red area denotes the time interval before the implementation of activity reduction. The green area denotes the $15$-days in which $\alpha$ decreases linearly from $1$ to $\alpha_\text{low}=0.176$. In the white area, activity is reduced to $\alpha_\text{low}$. The black, dash-dotted vertical line denotes the implementation of mobility restrictions. Orange curves illustrate the predictions from the model, blue curves are actual deaths data from \cite{protezioneCivile}.}
\label{fig:calibration}
\end{figure}

The parameter identification was formulated as a minimisation problem, solved by means of a dual-annealing procedure \cite{xiang2000efficiency}. Specifically, we defined the cost function $c$ as the weighted sum of the squared error between the number of deaths predicted by the model and the Regional real data, normalised with respect to the maximum number of deaths in the Region. To this aim, we defined the set of Regions $\mathcal R$ and the partition of Provinces into Regions as the function $\pi:\mathcal H\to\mathcal R$, such that $\pi(h)=r$ if and only if Province $h$ was located in Region $r$.  For each $r\in\mathcal R$, we introduced:
\begin{equation}
c^r=\sum_{t=t_0}^{t_{\text{end}}} 
\left(\frac{ D_{MA}^r (t)} {\bar D^r_{MA}} - 
\frac{ \sum_{h:\pi(h)=r} f R^h(t)}{ \bar D^r_{MA}}
\right)^2,
\end{equation}
and the cost function as the sum of $c^r$ weighted with the total number of deaths in the Region $r$
\begin{equation}
c=\sum_{r\in\mathcal R} \left( c^r \sum_{t=t_0}^{t_{\text{end}}}D^r(t) \right).
\end{equation}
Here, $D^r(t)$ indicates the reported deaths in Region $r$, $D^r_{MA}(t)$ the weekly moving average and $\bar D^r_{MA}=\max _t D^r_{MA}(t)$ its maximum value. Using the fatality ratio, the model predicts $f R^h(t)$ deaths in the Province $h$ at time $t$. Fig.~\ref{fig:calibration} shows real and simulated time-series with the identified parameters.

\section{Results}

\subsection{Implementation of NPIs}

Here, we elucidate the role of NPIs in halting the spread of COVID-19. We aim at disentangling the contribution of the two most common kinds of interventions: reduction of individuals' activity, through lockdown or social distancing, and enforcement of mobility restrictions. We take as a reference the NPIs implemented in Italy (detailed in the Supplementary material, Sec.~S2) and identify the NPI-related parameters from available data. The enforcement of lockdown and social distancing policies, gradually enacted during a time-window of two weeks (from 5th March 5 to 20th March) are modelled through a linear decrease of the $\alpha$ parameter from $1$ to  $\alpha_{\text{low}} = 0.176$.  The effect of the nearly complete mobility restrictions between Provinces has been observed from 1st March in the northern macro-regions and from 7th March in the southern ones \cite{pepe2020covid}. We model these restrictions by setting the mobility parameter to $\beta_{\text{low}}=0$ on the corresponding dates. 

\begin{figure}
\centering
    \includegraphics[width= \textwidth]{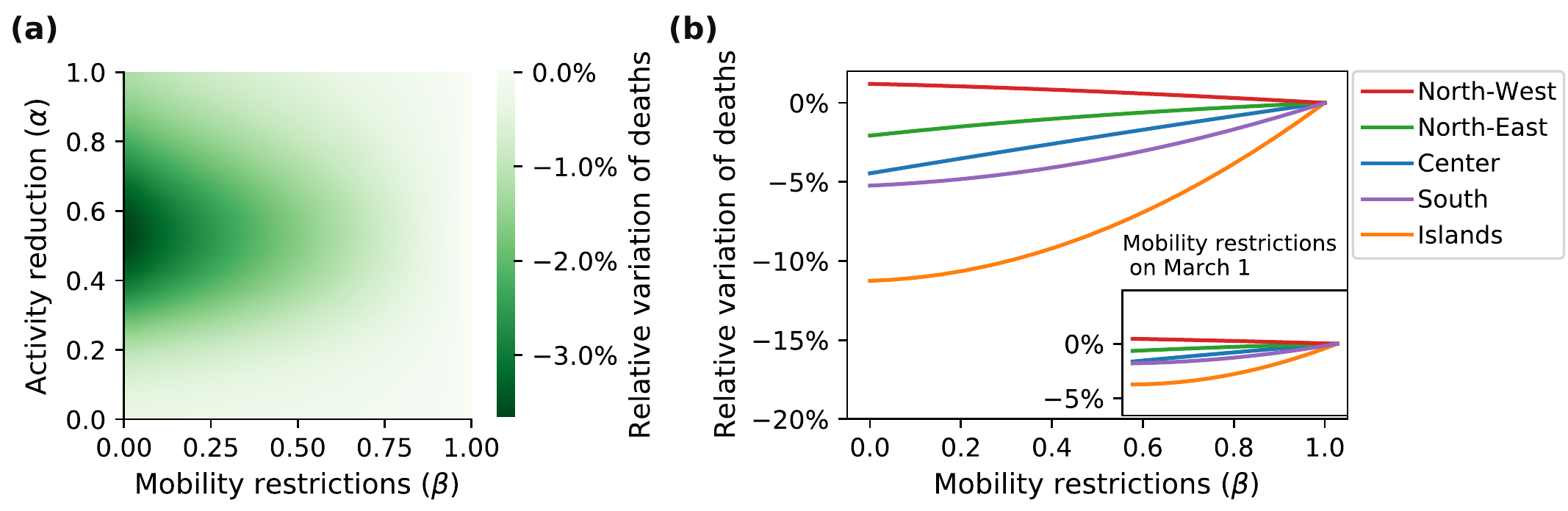}
\caption{Effect of early application of mobility restrictions. We consider the total number of deaths over a time-window of $104$ days from the beginning of the simulations (4th February) to the time corresponding to the relaxation of the most severe NPIs in Italy (18th May). 
In (a), we illustrate the interplay between the two NPI mechanisms, assuming an earlier application of mobility restrictions, and activity reduction applied at the original date (5th March). 
We consider different combinations of levels of activity reduction and mobility restrictions, {where higher levels of $\alpha$ or $\beta$ denote less severe NPIs}. The heat map codes the reduction of deaths with respect to a scenario where mobility restrictions are not applied.
Panel (b)  details the effect of earlier mobility restrictions at a macro-regional level, assuming a level of activity reduction as per the lockdown phase ($\alpha_{\text{low}} = 0.176$). The inset illustrates the effect corresponding to the application of the same restrictions at the actual date (1st March).}\label{fig:close_onlyBeta}
\end{figure}

We start from investigating the effect of mobility restrictions in combination with activity reduction policies (Fig.~\ref{fig:close_onlyBeta}a). 
We simulate the mobility restrictions as being applied on 4th February, that is, almost one month earlier than the actual date. 
We compare the number of deaths over the time-window that ranges from 4th February to the date of the first relaxation of NPIs in Italy (18th May). 
We observe that the effect of mobility restrictions becomes significant for intermediate levels of activity reduction policies (that is, $0.3<\alpha<0.7$). On the other hand, a negligible effect of mobility restrictions is registered for milder levels of activity reduction policies ($\alpha>0.7$) and for extremely severe activity reductions ($\alpha<0.3$). The latter, counter-intuitive finding, is due to a  balance between the increased number of deaths in some northern macro-regions (close to the initial outbreak) and the decrease of deaths in others (Supplementary material, Fig.~S5).

To detail this mechanism, we examine the number of deaths in each macro-region, using different levels of mobility restrictions and setting the activity reduction to the lockdown level, $\alpha_{\text{low}} = 0.176$ (Fig.~\ref{fig:close_onlyBeta}b).
Our results suggest that the impact of mobility restrictions is strongly dependent on their geographical location, and it vanishes if not timely implemented. Notably, we find that the  timely implementation of severe mobility restrictions would have reduced the number of deaths by more than $12\%$ in the \emph{Islands} macro-region (that is, far from where the outbreak was initially located) over the duration of severe NPIs. Such an advantage becomes smaller and smaller as the considered macro-regions are closer to the initial location of the outbreak. Paradoxically, mobility restrictions becomes even slightly detrimental if applied in the \emph{North-West} macro-region, where the outbreak started.  
This is due to the commute of infected individuals from the most affected Provinces to the rest of the Provinces and of susceptible individuals from less impacted Provinces to the rest of the country.  
For comparison, we also report death count for the implementation of the same restrictions on 1st March, corresponding to the actual date of their implementation.
We observe that the timing of NPIs is essential; an early application of travel restrictions by one month would have saved twice as many lives. Similar results are obtained for the peak of the epidemic incidence (Supplementary material, Figs.~S6 and~S7).

\begin{figure}
  \centering \includegraphics[width=\textwidth]{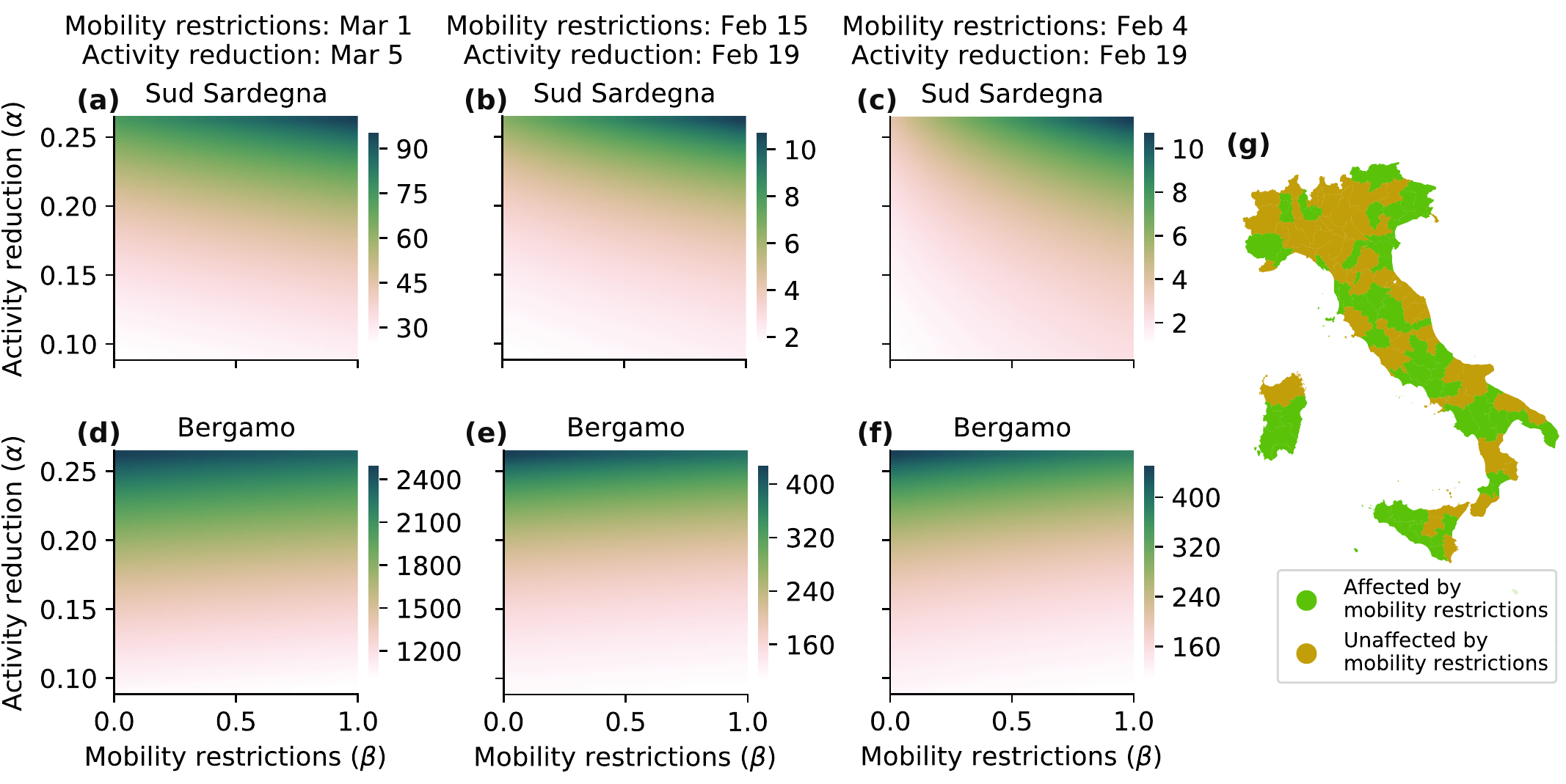}
  \caption{Effect of activity reduction and mobility restrictions, {where higher levels of $\alpha$ or $\beta$ denote less severe NPIs}.
  We consider the total number of deaths over a time-window of $104$ days from the beginning of the simulation (4th February) to the time corresponding to the relaxation of the most severe NPIs in Italy (18th May). We investigate three different intervention scenarios. In panels (a,d), both mobility restrictions and activity reduction are applied at the actual application dates. In panels (b,e) both strategies are hypothetically implemented earlier by $15$ days. In panels (c,g) mobility restrictions are further set earlier on 4th February, while keeping activity reduction applied earlier by only  15 days. 
  In panels (a--c), the province of \emph{Sud Sardegna} are illustrated as an example of the positive effect of early mobility restrictions. In panels (d--f), the province of \emph{Bergamo} look unaffected by mobility restrictions. 
  In (g), we illustrate the classification of Provinces in affected and unaffected by mobility restrictions, considering the scenarios relative to panels (c,f).}
  \label{fig:conponents}
\end{figure}

The large geographic variability of the number of deaths is confirmed in Figs.~\ref{fig:conponents}a--\ref{fig:conponents}f, which depict the total number of deaths for two {representative} Provinces under different timing and intensity of implementation of NPIs. While the Province of \emph{Bergamo} (in the \emph{North-West} macro-region), one of the earliest and biggest outbreaks, seems unrelieved by mobility restrictions, the Province of \emph{Sud Sardegna} (in the \emph{Islands} macro-region), an area much less affected by the pandemic than the former, would have largely benefited by such an intervention. 
To deepen this aspect, we factor out the role of the two types of NPIs by performing a non-negative matrix factorisation \cite{lee1999learning} on the outcome of our simulations at the Province level (details in the Supplementary material, Secs.~S3 and~S4).
Specifically, we focus on values of  $\alpha$ ranging over a $\pm 50\%$ interval with respect to the  value of the lowest activity coefficient $\alpha_{\text{low}}=0.176$, identified from real-world data during the lockdown, and we simulate the early application of mobility restrictions with different intensity levels and timing. {Our analysis leads to the characterisation of two sets of Italian Provinces}. The first (in green in Fig.~\ref{fig:conponents}{g}) comprises Provinces where timely implemented mobility restrictions are effective in reducing epidemic prevalence (for example, \emph{Sud Sardegna}). The second set (in brown in Fig.~\ref{fig:conponents}{g}) contains Provinces for which mobility restrictions have instead a negligible impact. Predictably, most of the Provinces in the \emph{North-West} (where the outbreak started) are unaffected by mobility restrictions, while the majority of Provinces in \emph{South} and \emph{Islands} would benefit from an early implementation of such restrictions. 

{Surprisingly, some important exceptions are identified}. For instance, the Provinces of \emph{Varese} and \emph{Monza} (close to the Milan metropolitan area) would have benefited from timely mobility restrictions. We believe that this is due to the initially small number of cases in those two Provinces, and to the large number of daily commuters from those Provinces to the \emph{Milan} Province and other neighbouring locations, where the Italian outbreak started. Hence, the same dynamics between North and South Italy is documented again over a much smaller spatial scale, between Northern Provinces with larger initial difference in epidemic prevalence. {Similar results are observed for other intervention scenarios} (Supplementary material, Fig.~S2).

\begin{figure}
\centering
\includegraphics[width=\textwidth]{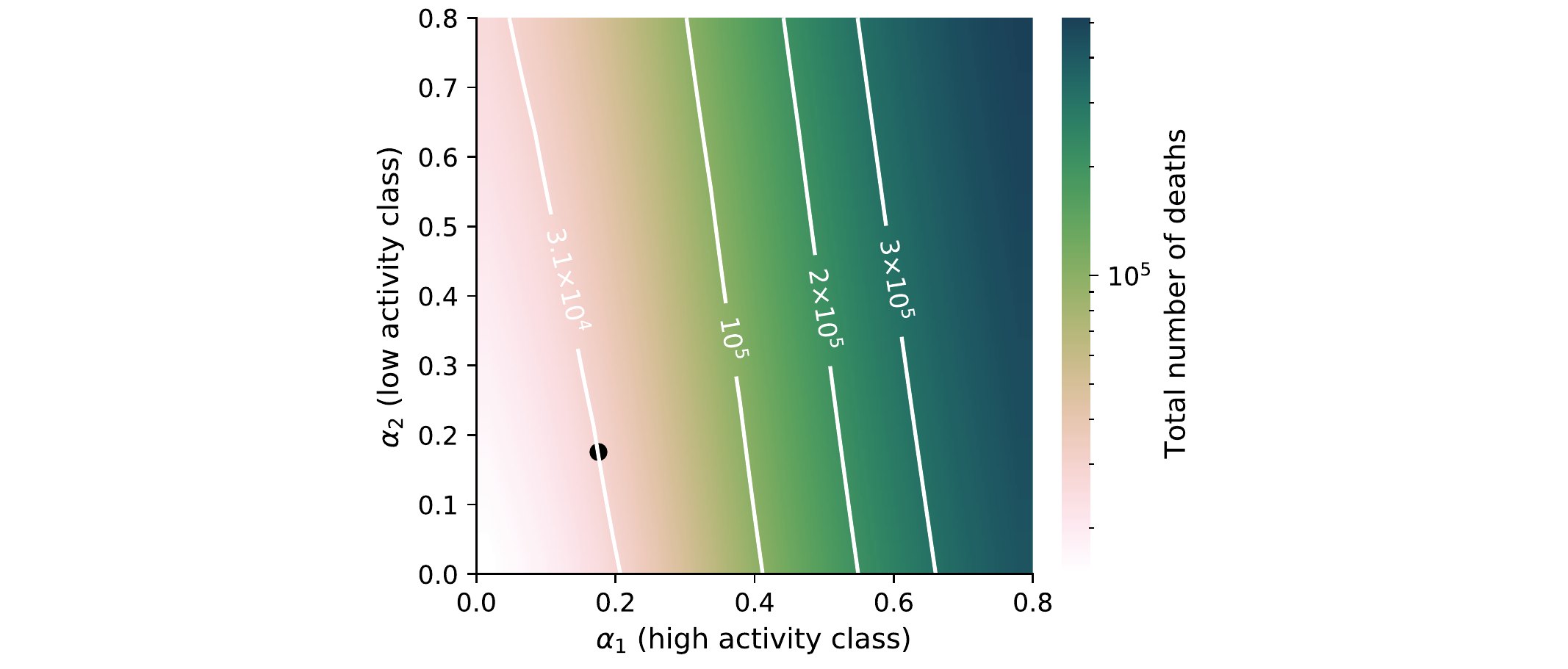}
\caption{Effect of targeted lockdown strategies. We consider the total number of deaths over a time-window of $104$ days from the beginning of the simulation (4th February) to the time corresponding to the relaxation of the most severe NPIs in Italy (18th May). On the two axes, $\alpha_1$ and $\alpha_2$ correspond to activity reduction of high and low activity classes, respectively, {where higher levels of $\alpha$ denote less severe NPIs}. Level curves are shown to clarify how targeted interventions should be combined to produce the same effect on the total number of deaths. The black dot represents the identified activity reduction in model calibration.
}\label{fig:lockMirati}
\end{figure}

Finally, we discuss the possibility of implementing targeted activity reductions that act independently on the two activity classes. This allows to study the effectiveness of differential intervention policies that could aim at strongly reducing social activity for age cohorts that are more at risk of developing severe illness, while implementing mild restrictions for younger people. Instead of a single parameter $\alpha$, we thus introduce two parameters $\alpha_1$ and $\alpha_2$ that measure the activity reduction for the high and the low activity class, respectively. The heat-map in Fig.~\ref{fig:lockMirati} illustrates the effect of different combinations of $\alpha_1$ and $\alpha_2$ on the total number of deaths; the level curves help understand the trade-off in targeting the two classes. We observe that the total number of deaths is mostly determined by the parameter $\alpha_1$, that is, the activity reduction for the high activity class. Hence, our results suggest that implementing targeted stay-at-home policies in which severe activity reductions are only enforced on the age cohorts that are more at risk (in our scenario, people over 65 years old) is not sufficient to reduce the overall death toll.

\subsection{Relaxation of NPIs toward reopening strategies}

The proposed meta-population model enables the analysis  of  reopening strategies to {relax}  restrictions while avoiding resurgent outbreaks. This has recently emerged as a key issue in the control of COVID-19 outbreaks in the medium- to long-term period \cite{ruktanonchai2020assessing}. We run our calibrated model to simulate the epidemic until the date of intervention {relaxation}. Then, we vary the values of   parameters $\alpha$ and $\beta$ to  account for the {relaxation} of the containment measures. Similar to the previous analysis, we consider a set of different options for the parameters after intervention {relaxation} and different times for starting the reopening strategies. Specifically, we model the {relaxation} of the reduction of social activity by varying the parameter $\alpha$ from $\alpha_{\mathrm{low}}$, identified during the lockdown, to a value $\alpha=0.6$. Likewise, we describe the uplifting of mobility restrictions by varying the parameter $\beta$ from $0$ (no mobility allowed) to $1$ (nominal mobility reinstated). 

\begin{figure}
  \centering \includegraphics[width=\textwidth]{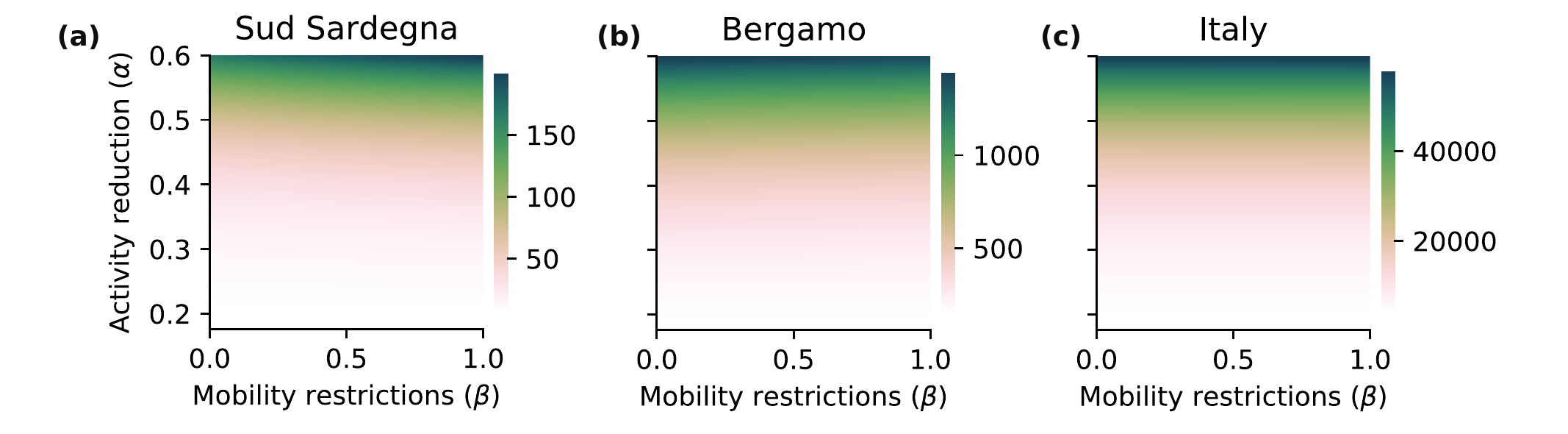}
  \caption{Effect of the relaxation of NPIs, for different levels of post-relaxation activity reduction and mobility restrictions, {where higher levels of $\alpha$ or $\beta$ denote less severe NPIs}. We consider the number of deaths over a time-window of $60$ days from the relaxation date (18th May). 
  In (a), we report the results for \emph{Sud Sardegna}, in (b) for \emph{Bergamo}, while in (c), we show the results aggregated at the country level.}
  \label{fig:conponents_open}
\end{figure}

Our results suggest that the effect of {maintaining mobility restrictions after} the relaxation of NPIs is negligible and dominated by the activity reduction (Fig.~\ref{fig:conponents_open}). We evaluate the total number of deaths in a time-window of $60$ days after the relaxation date (18th May). Both at the Province level, for which we show the examples of \emph{Sud Sargegna} (Fig.~\ref{fig:conponents_open}{a}) and \emph{Bergamo} (Fig.~\ref{fig:conponents_open}{b}), and at the aggregated country level (Fig.~\ref{fig:conponents_open}{c}), the contribution of mobility restrictions is little or absent. These results are confirmed by other scenarios with different relaxation dates (Supplementary material, Secs.~~S5 and~S6 and Figs.~S3 and~S4). 
Overall, this evidence indicates that activity reduction in the relaxation of NPIs should be thoughtfully calibrated, trading-off the risk of resurgent outbreaks and the social and economical costs associated with such policies. On the other hand, the further enforcement of mobility restrictions within the country {after the end of the epidemic wave} does not seem to be beneficial in the relaxation phase.

\section{Discussion}
Motivated by the evidence of the key role of NPIs in the ongoing COVID-19 outbreak \cite{prem2020effect, lai2020effect, kraemer2020effect,  tian2020, Haug2020}, we made an effort to propose a parsimonious mathematical framework to study NPIs and elucidate their impact on epidemic spreading. Specifically, we combined a meta-population model, capturing the spatial distribution of the population and its mobility patterns \cite{Pastor-Satorras2015,Gomes2018}, with an ADN-based structure, which reflects real-world features of social activity such as heterogeneity \cite{Perra2012,Zino2016} and behavioural traits \cite{Funk2010,RizzoPRE2014}. We explicitly incorporated two types of NPIs: actions aiming at reducing individuals' activity (social distancing, forbidding  gatherings and, in general, any  measure that curtails the number of contacts favouring the spread of the infection) and policies to restrict individuals' mobility (for instance, through travel bans). 
Through the lens of our modelling framework, we disentangled the effect of these two types of policies depending on the time of their implementation.
We calibrated the model with data on the ongoing COVID-19 outbreak in Italy \cite{protezioneCivile}.

We leveraged the model to explore a wide range of what/if scenarios on spatio-temporal dynamics of COVID-19 spreading for different combinations of NPIs. Our analysis allows to draw interesting conclusions on when and how to apply NPIs to make the fight against the spread more effective. While the level of activity reduction is unequivocally a decisive factor, the impact of mobility restrictions has a more {nuanced} impact. First, we observed that mobility restrictions produce benefits only if applied at the early stage of the outbreak, and only if paired with appropriate activity reduction policies. 
Moreover, we discovered that the effect of mobility restrictions is strongly dependent on space. In fact, through a non-negative matrix factorisation technique, we identified two sets of Provinces that are differently affected by mobility restrictions. The first set, mostly consisting of  Provinces in the North (where the outbreak initially started), has little or no benefit from mobility restrictions. The Provinces in the second set, instead, would have benefited from early implementation of mobility restrictions. Surprisingly, this set includes some of the Provinces in the north (most affected area).  Then, we discussed possible implementation of targeted NPIs, with severe restrictions only for age cohorts that are more at risk of developing severe illness. Our modelling framework brought to light concerning limitations in the implementation of these targeted interventions: although economical reasons may prompt these interventions, their public health value could be limited. Finally, while mobility restrictions are useful in the early stage of the outbreak, their late implementation is ineffective. A different scenario is observed for the relaxation of NPIs, where the level of activity reduction should be carefully and gradually {relaxed}.

Our study outlines several avenues of future research, which can be pursued leveraging the generality of the heterogeneous meta-population framework  proposed in this study.
{During the ``first wave" of COVID-19 in Italy,} NPIs have been homogeneously implemented nationwide through Decrees of the Prime Minister. {Hence, we have used uniform parameters among the Provinces. However, from November 2020, local NPIs have been enacted. The proposed} model could benefit from the study of heterogeneous
implementation (and relaxation) of NPIs between  Provinces and even the implementation of targeted mobility restrictions between specific Provinces (through the modification of the routing matrix $W$), whose analysis is envisaged for future research. The outcome of such an analysis can inform policymakers on targeted interventions that may reduce social and economical costs while effectively halting  the epidemic.  {Also, other targeted intervention policies, such as those leading to safe schools reopening, may be explored. These studies may be conducted at the entire country level or at a local level. Country-wise interventions could be engineered by utilising further activity classes that capture, for instance, students and teachers. In this vein, a contact matrix among activity classes could help capture inhomogeneous interaction patterns between and within activity classes, similar to \cite{mossong2008social}. Local targeted interventions, instead, may be pursued via our meta-population structure, where communities are used to model specific locations, such as schools and neighbourhoods.} The introduction of a community that represents the rest of the world would enable the study of the impact of closures of national borders. 

While we considered a simple model for the epidemic progression, additional compartments and transitions may be added to capture hospitalisation or testing \cite{giordano2020modelling}, and used to analyse different what/if scenarios {and design feedback control interventions, informed by the number of reported cases or hospitalisations \cite{dellarossa2020}. A limitation of our modelling framework lies in its deterministic formulation, which prevents it from capturing phenomena such as local  disease eradication. This may be crucial to study the pandemic at longer time-scales, encompassing future vaccination campaigns. A stochastic formulation of the meta-population activity-driven model may be proposed and utilised as a viable tool to shed light into these important phenomena and understand their impact on the spreading process. }
Finally, the simplicity of our mathematical framework may be conducive to a rigorous analytical treatment, involving for example the computation of the epidemic threshold, toward providing further insight into the effect of NPIs on the epidemic spreading.

\subsection*{Data availability} 
Epidemiological data are available at  \cite{protezioneCivile}; geographic, mobility and census data  at \cite{Istat}.

\subsection*{Code availability} 
The code used for the simulations is available at \url{https://gitlab.com/PoliToComplexSystemLab/adn-metapopulation-2020}.

\subsection*{Acknowledgments}
The authors are indebted to Alessandro Vespignani for precious discussion. 

\subsection*{Funding}
This work was partially supported by National Science Foundation (CMMI-1561134 and CMMI-2027990), Compagnia di San Paolo, MAECI (``Mac2Mic''), the European Research Council (ERC-CoG-771687), and the Netherlands Organisation for Scientific Research (NWO-vidi-14134).

\subsection*{Author Contributions}
L.Z. and A.R. conceived and designed the research with inputs from all the authors. F.P. performed the parameter identification and the numerical studies and wrote a first draft of the manuscript. A.R. and M.P. supervised the research and consolidated the manuscript in its present submission. All the authors contributed to the interpretation and analysis of the results and to reviewing the current submission of the manuscript.
\subsection*{Competing Interests} The authors declare that they have no competing interests.

\cleardoublepage

\includepdf[pages=-]{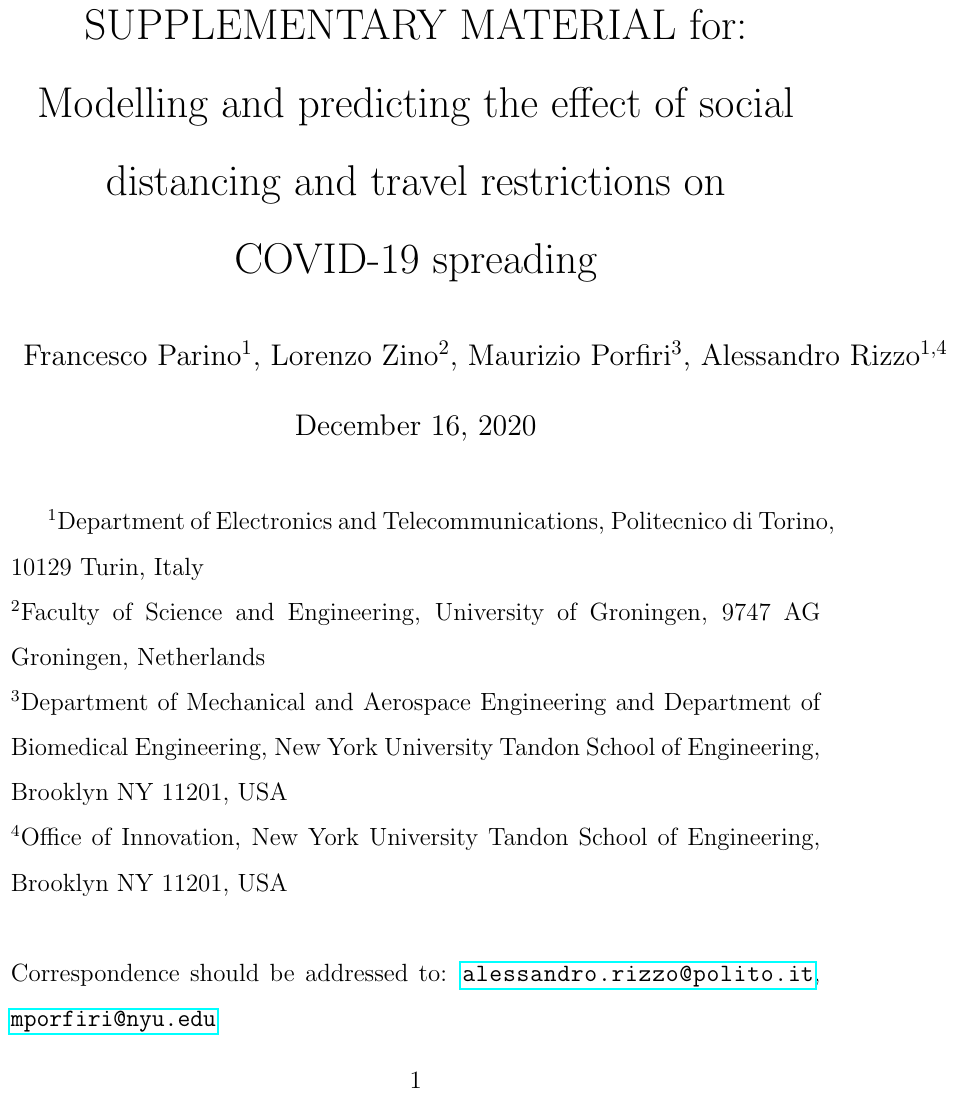}


\begin{thebibliography}{48}

\bibitem{whoSituation}
{World Health Organization}. {Coronavirus Disease ({COVID}-2019) Situation
  Reports}; 2020.
\newblock Accessed: \today.
\newblock Available at
  \url{www.who.int/emergencies/diseases/novel-coronavirus-2019/situation-reports}.

\bibitem{prem2020effect}
Prem K, Liu Y, Russell TW, Kucharski AJ, Eggo RM, Davies N, et~al.
\newblock {The effect of control strategies to reduce social mixing on outcomes
  of the {COVID}-19 epidemic in {Wuhan}, {China}: a modelling study}.
\newblock Lancet Public Health. 2020;5(5):e261--e270.
\newblock doi:10.1016/S2468-2667(20)30073-6.

\bibitem{lai2020effect}
Lai S, Ruktanonchai NW, Zhou L, Prosper O, Luo W, Floyd JR, et~al.
\newblock {Effect of non-pharmaceutical interventions to contain {COVID}-19 in
  {China}}.
\newblock Nature. 2020;585:410–413.
\newblock doi:10.1038/s41586-020-2293-x.

\bibitem{kraemer2020effect}
Kraemer MU, Yang CH, Gutierrez B, Wu CH, Klein B, Pigott DM, et~al.
\newblock The effect of human mobility and control measures on the {COVID}-19
  epidemic in {China}.
\newblock Science. 2020;368(6490):493--497.
\newblock doi:10.1126/science.abb4218.

\bibitem{tian2020}
Tian H, Liu Y, Li Y, Wu CH, Chen B, Kraemer MUG, et~al.
\newblock An investigation of transmission control measures during the first 50
  days of the {COVID}-19 epidemic in {China}.
\newblock Science. 2020;368(6491):638--642.
\newblock doi:10.1126/science.abb6105.

\bibitem{Haug2020}
Haug N, Geyrhofer L, Londei A, Dervic E, Desvars-Larrive A, Loreto V, et~al.
\newblock Ranking the effectiveness of worldwide {COVID-19} government
  interventions.
\newblock Nature Human Behaviour. 2020 Dec;4(12):1303--1312.
\newblock doi:10.1038/s41562-020-01009-0.

\bibitem{chinazzi2020effect}
Chinazzi M, Davis JT, Ajelli M, Gioannini C, Litvinova M, Merler S, et~al.
\newblock The effect of travel restrictions on the spread of the 2019 novel
  coronavirus ({COVID}-19) outbreak.
\newblock Science. 2020;368(6489):395--400.
\newblock doi:10.1126/science.aba9757.

\bibitem{Bartik2020economic}
Bartik AW, Bertrand M, Cullen Z, Glaeser EL, Luca M, Stanton C.
\newblock The impact of {COVID}-19 on small business outcomes and expectations.
\newblock Proc Natl Acad Sci USA. 2020;117(30):17656--17666.
\newblock doi:10.1073/pnas.2006991117.

\bibitem{Bonaccorsi2020economic}
Bonaccorsi G, Pierri F, Cinelli M, Flori A, Galeazzi A, Porcelli F, et~al.
\newblock Economic and social consequences of human mobility restrictions under
  {COVID}-19.
\newblock Proc Natl Acad Sci USA. 2020;117(27):15530--15535.
\newblock doi:10.1073/pnas.2007658117.

\bibitem{Qiu2020}
Qiu J, Shen B, Zhao M, Wang Z, Xie B, Xu Y.
\newblock A nationwide survey of psychological distress among {Chinese} people
  in the {COVID}-19 epidemic: implications and policy recommendations.
\newblock Gen Psychiatr. 2020 Mar;33(2):e100213--e100213.
\newblock doi:/10.1136/gpsych-2020-100213.

\bibitem{siegenfeld2020opinion}
Siegenfeld AF, Taleb NN, Bar-Yam Y.
\newblock Opinion: What models can and cannot tell us about {COVID}-19.
\newblock Proc Natl Acad Sci USA. 2020;117(28):16092--16095.
\newblock doi:10.1073/pnas.2011542117.

\bibitem{bertozzi2020challenges}
Bertozzi AL, Franco E, Mohler G, Short MB, Sledge D.
\newblock The challenges of modeling and forecasting the spread of {COVID}-19.
\newblock Proc Natl Acad Sci USA. 2020;117(29):16732--16738.
\newblock doi:10.1073/pnas.2006520117.

\bibitem{gatto2020spread}
Gatto M, Bertuzzo E, Mari L, Miccoli S, Carraro L, Casagrandi R, et~al.
\newblock Spread and dynamics of the {COVID}-19 epidemic in {Italy:} Effects of
  emergency containment measures.
\newblock Proc Natl Acad Sci USA. 2020;117(19):10484--10491.
\newblock doi:10.1073/pnas.2004978117.

\bibitem{aleta2020}
Aleta A, Mart{\'{\i}}n-Corral D, y~Piontti AP, Ajelli M, Litvinova M, Chinazzi
  M, et~al.
\newblock Modelling the impact of testing, contact tracing and household
  quarantine on second waves of {COVID}-19.
\newblock Nat Hum Behav. 2020;4(9):964--971.
\newblock doi:10.1038/s41562-020-0931-9.

\bibitem{metcalf2020mathematical}
Metcalf CJE, Morris DH, Park SW.
\newblock Mathematical models to guide pandemic response.
\newblock Science. 2020;369(6502):368--369.
\newblock doi:10.1126/science.abd1668.

\bibitem{modelingCovid19}
Vespignani A, Tian H, Dye C, Lloyd-Smith JO, Eggo RM, Shrestha M, et~al.
\newblock {Modelling {COVID}-19}.
\newblock Nat Rev Phys. 2020 Jun;2(6):279--281.
\newblock doi:10.1038/s42254-020-0178-4.

\bibitem{estrada2020covid}
Estrada E.
\newblock {{COVID}-19 and SARS-CoV-2. Modeling the present, looking at the
  future}.
\newblock Phys Rep. 2020;869:1--51.
\newblock doi:10.1016/j.physrep.2020.07.005.

\bibitem{dellarossa2020}
Della~Rossa F, Salzano D, Di~Meglio A, De~Lellis F, Coraggio M, Calabrese C,
  et~al.
\newblock A network model of Italy shows that intermittent regional strategies
  can alleviate the COVID-19 epidemic.
\newblock Nat Comm. 2020;11(1):5106.
\newblock doi:10.1038/s41467-020-18827-5.

\bibitem{colizza2006role}
Colizza V, Barrat A, Barth{\'e}lemy M, Vespignani A.
\newblock The role of the airline transportation network in the prediction and
  predictability of global epidemics.
\newblock Proc Natl Acad Sci USA. 2006;103(7):2015--2020.
\newblock doi:10.1073/pnas.0510525103.

\bibitem{brockmann2013hidden}
Brockmann D, Helbing D.
\newblock The hidden geometry of complex, network-driven contagion phenomena.
\newblock Science. 2013;342(6164):1337--1342.
\newblock doi:10.1126/science.1245200.

\bibitem{balcan2009multiscale}
Balcan D, Colizza V, Gon{\c{c}}alves B, Hu H, Ramasco JJ, Vespignani A.
\newblock Multiscale mobility networks and the spatial spreading of infectious
  diseases.
\newblock Proc Natl Acad Sci USA. 2009;106(51):21484--21489.
\newblock doi:10.1073/pnas.0906910106.

\bibitem{jia2020population}
Jia JS, Lu X, Yuan Y, Xu G, Jia J, Christakis NA.
\newblock {Population flow drives spatio-temporal distribution of {COVID}-19 in
  {China}}.
\newblock Nature. 2020 Jun;582(7812):389--394.
\newblock doi:10.1038/s41586-020-2284-y.

\bibitem{Pastor-Satorras2015}
Pastor-Satorras R, Castellano C, Van~Mieghem P, Vespignani A.
\newblock Epidemic processes in complex networks.
\newblock Rev Mod Phys. 2015;87:925--979.
\newblock doi:10.1103/RevModPhys.87.925.

\bibitem{Gomes2018}
G{\'o}mez-Garde{\~{n}}es J, Soriano-Pa{\~{n}}os D, Arenas A.
\newblock Critical regimes driven by recurrent mobility patterns of
  reaction--diffusion processes in networks.
\newblock Nat Phys. 2018;14(4):391--395.
\newblock doi:10.1038/s41567-017-0022-7.

\bibitem{Volz2009}
Volz E, Meyers LA.
\newblock Epidemic thresholds in dynamic contact networks.
\newblock J Royal Soc Interface. 2009;6:233--241.
\newblock doi:10.1098/rsif.2008.0218.

\bibitem{Holme2012}
Holme P, Saram\"{a}ki J.
\newblock {Temporal networks}.
\newblock Phys Rep. 2012;519:97--125.
\newblock doi:10.1016/j.physrep.2012.03.001.

\bibitem{Funk2010}
Funk S, Salath\'{e} M, Jansen VA.
\newblock {Modelling the influence of human behaviour on the spread of
  infectious diseases: a review}.
\newblock J R Soc Interface. 2010;7(50):1247--1256.
\newblock doi:10.1098/rsif.2010.0142.

\bibitem{RizzoPRE2014}
Rizzo A, Frasca M, Porfiri M.
\newblock Effect of individual behavior on epidemic spreading in activity
  driven networks.
\newblock Phys Rev E. 2014;90.
\newblock doi:10.1103/PhysRevE.90.042801.

\bibitem{Perra2012}
Perra N, Gon\c{c}alves B, Pastor-Satorras R, Vespignani A.
\newblock {Activity driven modeling of time varying networks}.
\newblock Sci Rep. 2012;2.
\newblock doi:10.1038/srep00469.

\bibitem{Zino2016}
Zino L, Rizzo A, Porfiri M.
\newblock {Continuous-time discrete-distribution theory for activity-driven
  networks}.
\newblock Phys Rev Lett. 2016;117.
\newblock doi:10.1103/PhysRevLett.117.228302.

\bibitem{brauer2011mathematical}
Brauer F, Castillo-Chavez C.
\newblock Mathematical models in population biology and epidemiology.
\newblock 2nd ed. New York NY, USA: Springer; 2012.
\newblock doi:10.1007/978-1-4614-1686-9.

\bibitem{protezioneCivile}
{Dipartimento della Protezione Civile}. {{COVID}-19 Italia - Monitoraggio
  situazione}. GitHub; 2020.
\newblock Accessed: \today.
\newblock Available at \url{https://github.com/pcm-dpc/COVID-19}.

\bibitem{giordano2020modelling}
Giordano G, Blanchini F, Bruno R, Colaneri P, Di~Filippo A, Di~Matteo A, et~al.
\newblock {Modelling the {COVID}-19 epidemic and implementation of
  population-wide interventions in Italy}.
\newblock Nat Med. 2020 Jun;26(6):855--860.
\newblock doi:10.1038/s41591-020-0883-7.

\bibitem{Kurtz1970}
Kurtz TG.
\newblock Solutions of Ordinary Differential Equations as Limits of Pure Jump
  {Markov} Processes.
\newblock J Appl Probab. 1970;7(1):49--58.
\newblock doi:10.2307/3212147.

\bibitem{Kurtz1971}
Kurtz TG.
\newblock Limit Theorems for Sequences of Jump {Markov} Processes Approximating
  Ordinary Differential Processes.
\newblock J Appl Probab. 1971;8(2):344--356.
\newblock doi:10.2307/3211904.

\bibitem{zhang2020evolving}
Zhang J, Litvinova M, Wang W, Wang Y, Deng X, Chen X, et~al.
\newblock Evolving epidemiology and transmission dynamics of coronavirus
  disease 2019 outside {Hubei} province, {China}: a descriptive and modelling
  study.
\newblock Lancet Inf Dis. 2020 Jul;20(7):793--802.
\newblock doi:10.1016/S1473-3099(20)30230-9.

\bibitem{world2020report}
{World Health Organization}. Report of the {WHO-China} joint mission on
  coronavirus disease 2019 ({COVID}-19). Geneva; 2020.
\newblock Accessed: \today.
\newblock Available at
  \url{https://www.who.int/publications/i/item/report-of-the-who-china-joint-mission-on-coronavirus-disease-2019-(covid-19)}.

\bibitem{zou2020sars}
Zou L, Ruan F, Huang M, Liang L, Huang H, Hong Z, et~al.
\newblock {SARS-CoV-2} viral load in upper respiratory specimens of infected
  patients.
\newblock N Engl J Med. 2020;382(12):1177--1179.
\newblock doi:10.1056/NEJMc2001737.

\bibitem{Istat}
ISTAT. {Istituto Nazionale di Statistica}; 2020.
\newblock Accessed: \today.
\newblock Available at \url{https://www.istat.it}.

\bibitem{mossong2008social}
Mossong J, Hens N, Jit M, Beutels P, Auranen K, Mikolajczyk R, et~al.
\newblock Social contacts and mixing patterns relevant to the spread of
  infectious diseases.
\newblock PLOS Med. 2008;5(3).
\newblock doi:10.1371/journal.pmed.0050074.

\bibitem{PEREZSAEZ2020}
Perez-Saez J, Lauer SA, Kaiser L, Regard S, Delaporte E, Guessous I, et~al.
\newblock {Serology-informed estimates of SARS-CoV-2 infection fatality risk in
  Geneva, Switzerland}.
\newblock Lancet Inf Dis. 2020.
\newblock doi:10.1016/S1473-3099(20)30584-3.

\bibitem{pepe2020covid}
Pepe E, Bajardi P, Gauvin L, Privitera F, Lake B, Cattuto C, et~al.
\newblock {COVID}-19 outbreak response, a dataset to assess mobility changes in
  {Italy} following national lockdown.
\newblock Sci Data. 2020 Jul;7(1):230.
\newblock doi:10.1038/s41597-020-00575-2.

\bibitem{backer2020incubation}
Backer JA, Klinkenberg D, Wallinga J.
\newblock Incubation period of 2019 novel coronavirus ({2019-nCoV}) infections
  among travellers from {Wuhan}, {China}, 20--28 January 2020.
\newblock Eurosurveillance. 2020;25(5):2000062.
\newblock doi:10.2807/1560-7917.ES.2020.25.5.2000062.

\bibitem{woelfel2020clinical}
W{\"o}lfel R, Corman VM, Guggemos W, Seilmaier M, Zange S, M{\"u}ller MA,
  et~al.
\newblock Virological assessment of hospitalized patients with {COVID-2019}.
\newblock Nature. 2020 May;581(7809):465--469.
\newblock doi:10.1038/s41586-020-2196-x.

\bibitem{Linton2020}
Linton N, Kobayashi T, Yang Y, Hayashi K, Akhmetzhanov A, Jung Sm, et~al.
\newblock Incubation Period and Other Epidemiological Characteristics of 2019
  Novel Coronavirus Infections with Right Truncation: A Statistical Analysis of
  Publicly Available Case Data.
\newblock J Clin Med. 2020 Feb;9(2):538.
\newblock doi:10.3390/jcm9020538.

\bibitem{xiang2000efficiency}
Xiang Y, Gong X.
\newblock Efficiency of generalized simulated annealing.
\newblock Phys Rev E. 2000;62(3):4473.
\newblock doi:10.1103/physreve.62.4473.

\bibitem{lee1999learning}
Lee DD, Seung HS.
\newblock Learning the parts of objects by non-negative matrix factorization.
\newblock Nature. 1999;401(6755):788--791.
\newblock doi:10.1038/44565.

\bibitem{ruktanonchai2020assessing}
Ruktanonchai NW, Floyd JR, Lai S, Ruktanonchai CW, Sadilek A, Rente-Lourenco P,
  et~al.
\newblock {Assessing the impact of coordinated {COVID}-19 exit strategies
  across Europe}.
\newblock Science. 2020;369:1465--1470.
\newblock doi:10.1126/science.abc5096.

\end{thebibliography}
\end{document}